\documentclass[10pt,conference]{IEEEtran}

\usepackage[T1]{fontenc}
\usepackage[utf8]{inputenc}
\usepackage{amsmath,amssymb}
\usepackage{graphicx}
\usepackage{booktabs}
\usepackage{array}
\usepackage{tabularx}
\usepackage{tikz}
\usetikzlibrary{arrows.meta,positioning,shapes.geometric}
\usepackage{cite}
\usepackage{url}
\usepackage{microtype}
\usepackage{balance}

\newcommand{\CASE}{\textsc{CASE}}
\newcommand{\valid}{\mathsf{valid}}
\newcommand{\invalid}{\mathsf{invalid}}
\newcommand{\unknown}{\mathsf{unknown}}

\title{When Relevant Code Is Wrong: Runtime-Aware Context for AI Coding Agents}

\author{\IEEEauthorblockN{Wesley K. G. Assunção}
\IEEEauthorblockA{\textit{North Carolina State University}\\
Raleigh, NC, USA}
\and
\IEEEauthorblockN{Ramesh Periyathambi}
\IEEEauthorblockA{\textit{eBay Inc.}\\
San Francisco, CA, USA}
}

\begin{document}
\maketitle

\begin{abstract}
AI coding agents increasingly rely on retrieved code and development history to guide software changes. However, highly relevant artifacts may describe implementations that are inactive or incompatible with the target deployment. Neither semantic similarity nor recency establishes whether retrieved information remains applicable under feature flags, architectural migrations, and deployment-specific configurations. We challenge relevance-oriented retrieval by introducing \emph{operational context validity}, a task-, deployment-, and time-dependent criterion for determining whether software evidence can legitimately guide implementation decisions. We instantiate this new retrieval principle in CASE, an agent-agnostic framework that connects repository artifacts with runtime configurations and architectural evolution, classifies evidence as \emph{valid}, \emph{invalid}, or \emph{unknown}, and generates context packages distinguishing actionable guidance from historical rationale and unresolved constraints. We investigate its feasibility through a reproducible preliminary evaluation involving six development tasks, 15 artifacts, and three deployments. Conventional retrieval exposes 16 incompatible artifact occurrences as actionable, whereas CASE excludes them while retaining all applicable artifacts under the prototype's reference rules. Recency filtering reduces valid recall to 0.43, while removing feature-flag evidence from CASE reduces it to 0.13. Although these results demonstrate feasibility rather than improvements in coding-agent correctness, they reveal limitations of relevance-based retrieval and motivate a new research direction: enabling coding agents to reason not only about which information is relevant, but also whether it remains operationally valid.
\end{abstract}


\section{Introduction}
\label{sec:introduction}

AI coding agents are increasingly capable of performing complex software engineering tasks~\cite{yang2024swe}. They can explore repositories, retrieve relevant historical pull requests (PRs), and propose changes across multiple files~\cite{yin2026chronos}. Existing research has demonstrated the importance of repository-level context for solving development tasks~\cite{jimenez2024swebench}, motivating techniques that retrieve relevant code, documentation, and software dependencies~\cite{gu2026retrieve,ouyang2025repograph}. However, these approaches predominantly focus on identifying information that seems relevant to a task, often overlooking a fundamental question: \emph{Is the retrieved information operationally valid for the task at hand?}

This distinction becomes particularly important in large, continuously evolving software systems, where multiple implementations, architectural versions, and runtime configurations coexist~\cite{lercher2024api,
rahman2016toggles,ayas2023migration}. Our research is motivated by software evolution challenges encountered in collaboration with a large online marketplace. 
In their systems, repositories and development artifacts often capture multiple stages of a system's evolution rather than a single, authoritative view of how it currently operates. Different implementations may coexist, services may have been replaced, and feature flags or deployment-specific configurations may determine which execution paths are active~\cite{ducasse2009reconstruction,weng2026stale}. Hence, source code, PRs, and documentation may seem relevant to a development task even though they describe implementations that are no longer applicable to the target environment. In this context, existing retrieval approaches can identify artifacts related to a task but do not necessarily establish whether their architectural assumptions, dependencies, and runtime conditions remain valid. Consequently, coding agents may propose technically plausible changes that are incompatible with the system's actual operational state.

The underlying problem is that \emph{relevant information is not necessarily applicable information}. Recent research has demonstrated that retrieved repository context can be temporally stale and induce obsolete API usage~\cite{weng2026stale}, that irrelevant or misleading retrieved information can harm code generation~\cite{wu2024repoformer}, and that repository-specific agent knowledge can become outdated as software evolves~\cite{duan2026repo2skill}. Evolution-aware approaches, such as Chronos, further demonstrate the value of retrieving historical development decisions~\cite{yin2026chronos}. However, these contributions do not directly address \emph{deployment-conditioned prescriptive validity}: determining whether retrieved evidence can legitimately guide implementation decisions under a particular runtime configuration and deployment state.


We argue that this limitation calls for a shift from \emph{relevance-based retrieval} toward \emph{operationally valid context selection}. We introduce \emph{operational context validity}, which characterizes whether a retrieved software artifact can legitimately guide an implementation decision for a particular deployment and point in time. Building on this idea, we propose \CASE{}, a context-aware framework for existing coding agents. Rather than simply retrieving additional information, \CASE{} connects software artifacts with runtime configurations, feature flags, and architectural history to distinguish applicable implementations from incompatible or uncertain evidence. It then generates task-specific context packages that identify which artifacts can guide implementation, which require further verification, and which should be retained only as explanatory history.

Our central hypothesis is that explicitly distinguishing relevance from operational validity can reduce implementation decisions that conflict with the target system's architecture and runtime configuration. We investigate this forward-looking idea through three contributions: (1) a formulation of \emph{operational context validity} as a deployment- and time-dependent property of retrieved software evidence; (2) a lightweight, validity-aware context selection mechanism that can be integrated with existing coding agents; and (3) an initial proof-of-concept pilot evaluation demonstrating the feasibility of deployment-aware context selection. These contributions establish a foundation for investigating whether operationally valid context can improve the reliability of AI-assisted software evolution.

\section{Operational Context Validity}
\label{sec:validity}

\begin{figure*}[t]
    \centering
    \includegraphics[width=.85\linewidth]{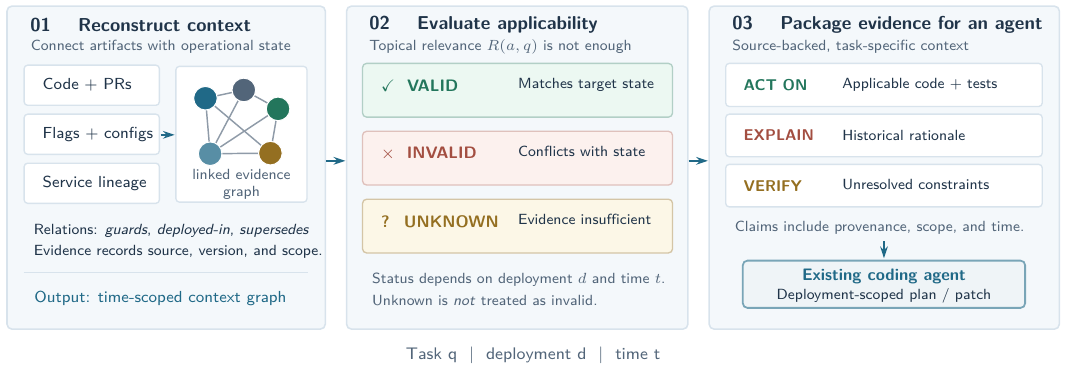}
    \caption{Overview of CASE. The framework reconstructs software context from heterogeneous artifacts, assesses their operational validity for a target deployment and system state, and generates a task-specific context package for an AI coding agent. Applicable implementations are distinguished from obsolete historical evidence and unresolved configuration constraints.}
    \label{fig:case}
\end{figure*}

Let $q$ denote a maintenance task, $t$ the target system snapshot, and $d$ a deployment context (e.g., geographic region with different regulations, enabled flags, service versions, and rollout cohort). Let $a$ be a candidate artifact (e.g., code, test, PR, architectural decision, or incident report). 

Conventional retrieval primarily estimates topical relevance $R(a,q)$, the relevance of an artifact $a$ to a task $q$. This relevance can be estimated through lexical matching (e.g., BM25~\cite{robertson2009}), semantic similarity~\cite{gu2026retrieve}, or graph-based relationships among repository artifacts~\cite{ouyang2025repograph}.
A high relevance score indicates that an artifact is closely related to the requested functionality and may provide useful implementation information. However, relevance alone does not establish whether the artifact reflects the system's current operational state. 

In this work, we define an artifact's \emph{prescriptive applicability} $V$ to deployment $d$ at time $t$:
\begin{equation}
\label{eq:validity}
V(a\mid d,t)=
\begin{cases}
\valid, & E_t\models \pi_a(d),\\
\invalid, & E_t\models \neg\pi_a(d),\\
\unknown, & \text{otherwise}.
\end{cases}
\end{equation}
where $\pi_a$ expresses the preconditions under which the implementation represented by $a$ applies, and $E_t$ is timestamped evidence from deployment manifests, flag configurations, service routing, and migration records. A definitive label requires sufficiently authoritative, non-conflicting evidence; missing or contradictory information yields $\unknown$. Applicability is thus \emph{not} inferred merely from a lack of runtime traces, which may reflect sampling or observability gaps.
Applicability labels characterize whether an artifact can guide implementation decisions, not whether its historical content is factually correct. An operationally invalid artifact may therefore remain useful as explanatory evidence.

$V$ is deliberately separate from $R$. A migration PR might have high $R$ but $V=\invalid$ for implementing the current path. It can still be included as \emph{historical rationale}, explicitly labeled as such. Conversely, a less textually similar artifact can be the applicable implementation. This distinction prevents a binary ``old means irrelevant'' rule: artifact \emph{roles} (prescriptive versus explanatory) are part of the retrieval decision.

\noindent\textbf{Evidence and scope are distinct from age.} To derive $\pi_a$, a lightweight prototype can use explicit source-code guards, service deployment mappings, rollout rules, and migration/retirement records. A commit date is only a clue: a long-lived implementation may remain active, and a new commit can target a disabled flag. Sources also have different authority. For example, a versioned flag assignment can establish that a path is enabled for a specified cohort, whereas absence from sampled production logs alone cannot establish that it is disabled. Each applicability assertion should therefore identify its source, timestamp, scope, and any unresolved conflict.


A useful counterexample is a \emph{regional rollback}: the same old REST endpoint may be $\invalid$ for deployment $d_1$ yet $\valid$ for $d_2$ at the same time $t$. Therefore, neither globally deleting outdated-looking artifacts nor asking an LLM to prefer newer files implements operational validity. The intended inference is contextual, and the system must allow legitimate disagreement between deployments.

\noindent\textbf{Core hypothesis.} Holding model, candidate pool, and context-token budget constant, evidence packages that distinguish $\valid$, $\invalid$, and $\unknown$ artifacts will reduce deployment-incompatible implementation decisions compared with similarity-only retrieval. Gains should concentrate in systems with architectural migrations or configuration-dependent execution; little or no gain is expected for stable, single-configuration components.

\section{A Minimal \CASE{} Context Framework}
\label{sec:approach}
The longer-term vision of \CASE{} involves context reconstruction, runtime-aware historical retrieval, and task-specific context packages. In this preliminary work, our goal is to test the central idea with a preliminary evaluation. Figure~\ref{fig:case} presents an overview of our framework, described in what follows.


\noindent\textbf{Step 1: Reconstruct context.} Start with a top-$k$ repository/PR retrieval pool and create a small, versioned context graph for the requested service. Its nodes represent artifacts, implementations, flags, deployments, and services; its edges include \emph{guards}, \emph{deployed-in}, \emph{calls}, and \emph{supersedes}. Links may come from configuration exports, service inventories, PR metadata, and a limited set of developer-verified mappings. The graph is a mechanism for collecting validity evidence.

\noindent\textbf{Step 2: Evaluate applicability.} For each candidate artifact, evaluate $V(a\mid d,t)$ with explicit provenance. A retired REST path is excluded from \emph{prescriptive} recommendations if the target service has been superseded; a flag-guarded implementation is only considered applicable when its activation conditions are supported. Unknowns remain visible and may trigger a question or additional retrieval. Because flags can be conditional on user cohorts and live rollouts, a global ``on/off'' assumption is insufficient.

\noindent\textbf{Step 3: Package evidence for an agent.} Construct $P_B(q,d,t)$ under a token budget $B$, deployment $d$ at time $t$. Each actionable claim contains an artifact pointer, applicability scope, evidence timestamp, and status. The package prioritizes currently applicable patterns, retains obsolete decisions only as labeled explanatory history, and states important unresolved constraints and relevant tests. An existing agent receives the package through its normal context interface; no model fine-tuning or autonomous production modification is assumed.



\section{Preliminary Evaluation}
\label{sec:pilot}

We conducted a preliminary evaluation to investigate whether operational validity can improve the selection of software context and distinguish applicable implementation evidence from outdated or deployment-incompatible artifacts. We address three research questions (RQs): 
\begin{itemize}
    \item \textbf{RQ1:} To what extent does topical relevance diverge from operational validity?
    \item \textbf{RQ2:} Can validity-aware context selection reduce the inclusion of operationally incompatible artifacts?
    \item \textbf{RQ3:} How do runtime configurations and architectural history influence context selection?
\end{itemize}

\noindent\textbf{Prototype and Experimental Setup.}
We implemented a proof of concept for \CASE{} using an illustrative merchant-trust application developed in Java with Spring Boot. Although illustrative, this example mirrors a real case faced by a large online marketplace.
The prototype, including its source code, deployment configurations, evaluation tasks, and reproducible experimental results, is publicly available in our replication package~\cite{caseMerchantTrust2026}.
The application exposes a merchant trust score through two alternative implementations: a legacy REST endpoint and a newer GraphQL architecture supporting pluggable signals. The applicability of each implementation depends on deployment-specific feature flags and architectural evolution. For example, the U.S. deployment migrated to GraphQL, whereas the European deployment subsequently rolled back to REST. The Asia-Pacific deployment contains a partially enabled GraphQL configuration with unresolved cohort information.

The evaluation comprises 15 artifacts, including source code, tests, historical pull requests, and architectural documentation, across three deployments and six development tasks. The tasks cover architectural migrations (two), feature-flag and regional variations (two), and stable components (two). Each task specifies a target deployment and time, and only artifacts available at that time are considered. A BM25 retriever~\cite{robertson2009} produces the top-eight candidate artifacts for each task.

We compare four conditions: (1) \emph{Semantic Only}, which retrieves artifacts based on topical relevance; (2) \emph{Recency Filtered}, which applies a 180-day recency window; (3) \emph{Semantic + Raw Configuration}, which supplements retrieved artifacts with deployment configuration without assessing applicability; and (4) \emph{CASE Validity Package}, which classifies artifacts as valid, invalid, or unknown using timestamped configuration and architectural evidence. All conditions start from the same candidate pool and use a common budget of approximately 1,500 tokens. We additionally evaluate two ablations that remove feature-flag evidence or architectural lineage information.

We measure \emph{valid precision}, the proportion of actionable artifacts that are applicable to the target deployment; \emph{valid recall}, the proportion of applicable artifacts in the candidate pool retained as actionable; and the numbers of incompatible and unknown artifacts presented as actionable. For conventional retrieval, all retrieved artifacts are considered actionable. For CASE, only artifacts classified under \textsc{Act On} are actionable, while uncertain and incompatible artifacts are retained under \textsc{Verify} and \textsc{Explain}, respectively. Precision and recall are averaged across tasks.

\begin{table}[t]
\centering
\caption{Results of the synthetic pilot. Precision (P) and
recall (R) are task-level means; incompatible (Inc.) and
unknown (Unk.) are total actionable artifact occurrences.}
\label{tab}
\scriptsize
\setlength{\tabcolsep}{3.5pt}
\begin{tabular}{@{}lcccc@{}}
\toprule
\textbf{Condition} & \textbf{P} & \textbf{R} & \textbf{Inc.} & \textbf{Unk.} \\
\midrule
Semantic Only       & 0.50 & 1.00 & 16 & 5 \\
Recency Filtered    & 0.43 & 0.43 &  6 & 5 \\
Semantic + Raw Cfg. & 0.50 & 1.00 & 16 & 5 \\
\textbf{CASE}       & \textbf{1.00} & \textbf{1.00} & \textbf{0} & \textbf{0} \\
CASE w/o Flags      & 0.33 & 0.13 &  0 & 0 \\
CASE w/o Lineage    & 1.00 & 1.00 &  0 & 0 \\
\bottomrule
\end{tabular}
\end{table}

\noindent\textbf{Results.}
Table~\ref{tab} summarizes the deterministic replay results.

\emph{RQ1: Relevance does not guarantee operational validity.}
Although semantic retrieval identifies all applicable artifacts within the candidate pools (recall of 1.00), only half of the artifacts it presents as actionable are applicable on average (precision of 0.50). Across six tasks, it exposes 16 incompatible and five uncertain artifact occurrences. Recency filtering reduces incompatible occurrences to six but also decreases recall to 0.43. Notably, after the European deployment rolled back to REST, the four artifacts retained by recency filtering for that task were all incompatible. These observations illustrate why neither relevance nor recency alone establishes applicability.

\emph{RQ2: Explicit validity assessment changes actionable context.}
CASE achieves precision and recall of 1.00 in the synthetic replay, with no incompatible or uncertain artifacts presented as actionable. In contrast, supplementing semantic retrieval with raw configuration data produces the same artifact-selection metrics as semantic retrieval alone. This difference reflects CASE's explicit classification and separation of retrieved evidence, rather than the availability of additional configuration information. Whether a coding agent could independently infer applicability from raw configuration remains an open question.

\emph{RQ3: Runtime evidence determines context applicability.}
Removing feature-flag evidence reduces CASE's recall from 1.00 to 0.13, as previously actionable artifacts become uncertain. Removing architectural lineage evidence produces no change in this fixture because feature-flag information already determines the applicable paths. We additionally replayed identical development requests against different deployments while holding the repository snapshot constant. CASE correctly changes its actionable context from the GraphQL implementation in \texttt{us-prod} to the REST implementation in \texttt{eu-prod}, demonstrating deployment-dependent context selection.


\noindent\textbf{Limitations.} Our pilot is a feasibility evaluation rather than an independent assessment of effectiveness. Artifact relationships and applicability conditions are manually curated, and the reference labels follow the same rules implemented by CASE's evaluator. Consequently, its perfect precision and recall are expected under the fixture's assumptions and should not be interpreted as evidence of generalization. Furthermore, the evaluation measures context selection, not the correctness of coding-agent decisions or generated patches. Larger studies with independently established ground truth and real development tasks are necessary to determine whether validity-aware packages improve downstream agent performance.

\section{New Ideas and Emerging Results}

CASE introduces a new perspective on repository-aware retrieval: beyond identifying relevant artifacts, coding agents must know whether those artifacts can legitimately guide implementation decisions. The preliminary evaluation reveals two important implications. First, identical tasks can require different implementation evidence across deployments, challenging retrieval approaches that assume a single repository state. Second, feature-flag evidence substantially influences applicability decisions, whereas architectural-lineage information provides no additional benefit in the current synthetic fixture.

These emerging results motivate treating operational validity as an explicit, evidence-backed property of retrieved context rather than an implicit consequence of relevance or recency. The resulting research direction is to investigate whether validity-aware context selection can improve coding-agent decisions while accounting for incomplete evidence, evolving configurations, and the cost of context reconstruction.

\section{Related Work}
Retrieval-augmented code generation (CodeRAG) improves access to repository-specific information, but its effectiveness depends on which artifacts are selected: similar-code retrieval can introduce irrelevant or misleading context~\cite{gu2026retrieve,hai2025contexts}. Structural approaches such as DraCo, RepoGraph, and RepoHyper use dataflow or repository graphs to recover cross-file relationships and improve retrieval~\cite{cheng2024draco,ouyang2025repograph,phan2025repohyper}. More recently, Chronos has connected historical pull requests to help agents reason about software evolution and reuse engineering experience~\cite{yin2026chronos}. Behavioral studies of coding agents likewise motivate examining how agents use context rather than relying on resolution rates alone~\cite{mehtiyev2026behavior}. These approaches motivate richer contextual representations; CASE addresses a complementary decision: whether a retrieved artifact should be used as \emph{prescriptive evidence} for a particular system state.

Weng et al. show that stale cross-file snippets can induce obsolete API usage~\cite{weng2026stale}. However, operational validity is not equivalent to recency. Feature toggles can activate different paths for different deployments or user cohorts~\cite{rahman2016toggles}, and studies of their lifecycle show that toggles may persist or be removed as systems evolve~\cite{hoyos2021removal}. Research also connects feature toggles and configuration options within a common software-configuration model~\cite{mahdavihezaveh2024configuration}. CASE builds on these insights by separating topical relevance from deployment- and time-conditioned applicability.



\section{Future Plans}

We plan to extend CASE with provenance-preserving integration of configuration, runtime, and architectural evidence, and evaluate it using independently annotated development tasks. Controlled experiments will compare CASE with conventional retrieval, recency filtering, and simpler configuration-aware alternatives under equivalent context budgets. We will assess both context-selection quality and downstream outcomes, including implementation planning, patch correctness, and deployment compatibility. We will also investigate cases requiring architectural-lineage reasoning, incomplete or conflicting configuration evidence, and the trade-offs among context accuracy, abstention, retrieval cost, and scalability.
Ultimately, we aim to establish whether explicitly reasoning about the operational validity of software evidence can make AI coding agents more reliable in evolving, configurable software systems.

\section{Conclusion}

This work introduces \emph{operational context validity} as a fundamental requirement for AI coding agents: retrieved software evidence must not only be relevant to a development task but also applicable to its target deployment and system state. CASE operationalizes this principle by connecting repository artifacts with runtime configurations and architectural evolution, distinguishing actionable evidence from incompatible historical information and unresolved constraints. Our preliminary evaluation demonstrates the feasibility of this approach: in a synthetic merchant-trust scenario, CASE excluded 16 incompatible artifact occurrences that conventional retrieval exposed while retaining all applicable artifacts under the prototype's reference rules. 

\section*{Data Availability}
A replication package for our pilot is publicly available~\cite{caseMerchantTrust2026}.

\balance
\bibliographystyle{IEEEtran}
\bibliography{references}
\end{document}